\documentclass[twocolumn,english]{revtex4-1}
\usepackage[T1]{fontenc}
\usepackage[latin9]{inputenc}
\usepackage{amsmath}
\usepackage{graphicx}
\usepackage{xcolor}
\usepackage{babel}

\begin{document}
\title{Training overdamped dynamics}
\author{Marc Berneman}
\affiliation{Faculty of Mechanical Engineering, Technion, 320000 Haifa, Israel}
\author{Daniel Hexner}
\affiliation{Faculty of Mechanical Engineering, Technion, 320000 Haifa, Israel}
\begin{abstract}
In regimes where inertia is negligible, the temporal evolution is
governed by overdamped dynamics. This limit is particularly relevant
in soft-matter contexts, such as polymers, colloidal suspensions,
and processes occurring at the cellular scale. Being able to manipulate the dynamics of such many-particle systems would enable control over
rate-dependent elastic responses, time-dependent material properties,
relaxation processes, and perhaps the hydrodynamics of suspensions.
In this work, we develop a framework for manipulating overdamped dynamics
through local, physically motivated update rules. Our approach is
inspired by ideas from physical learning and directed aging, in which
microscopic parameters adapt autonomously to endow a material with
a desired function. Using the Rayleighian formulation, whose minimization
reproduces the overdamped equations of motion, we derive approximate
directed-aging and contrastive rules tailored to dissipative
systems. To demonstrate these ideas, we study a disordered Maxwell
material that behaves elastically at short times but flows at long
times. By locally modifying the viscous damping, we show that one
can tune the viscous Poisson's ratio and shape local mechanical responses.
These results illustrate how materials can be trained to exhibit targeted
rate-dependent elastic and viscous behaviors.
\end{abstract}
\maketitle

\section{Introduction}
Systems in which inertia is negligible are governed by overdamped dynamics, in which the velocity of each degree of freedom is determined by the internal and external forces acting on it. Such dynamics arise in a wide range of physical, chemical, and biological settings, including polymers~\cite{doi1988theory}, colloidal suspensions~\cite{brady1988stokesian}, cellular and tissue mechanics~\cite{farhadifar2007influence,bi2016motility}, reaction networks, and models of collective animal behavior~\cite{vicsek1995novel}. Despite the apparent simplicity of their component-level dynamics, overdamped systems can exhibit rich emergent collective behavior.

Controlling these dynamics could enable materials and complex systems with prescribed temporal evolution and rate-dependent functionality. Conventionally, the behavior of a complex system is designed by tuning its microscopic interaction parameters. Examples include local stiffnesses and geometry in elastic structures~\cite{sigmund1996composites,bertoldi2017flexible,goodrich2015principle,rocks2017designing,zu2025fully,arzash2025rigidity,alqatari2024epistatic}, interaction affinities in self-assembling systems~\cite{mirkin2020dna,glotzer2007anisotropy,mcmullen2018freely,sacanna2011shape}, and conductances in resistor networks~\cite{rocks2019limits}. This approach requires solving an inverse-design problem to identify the appropriate microscopic parameters and then physically implementing the resulting design.

An alternative approach is to use self-organization to modify material parameters, thereby controlling system-level properties without directly fabricating or manipulating the microstructure~\cite{pashine2019directed,kedia2023drive,ong2024jamming}. In directed aging~\cite{pashine2019directed,arinze2023learning,hagh2022transient,gowen2025training} and physical learning~\cite{scellier2017equilibrium,stern2021supervised,anisetti2023learning,stern2023learning,anisetti2024frequency,altman2024experimental,ezraty2025harnessing,eran2025multistable,arzash2025learning,du2025metamaterials,patil2023self}, microscopic interaction parameters evolve autonomously in response to external forcing. This local evolution may arise from natural physical processes, such as plasticity, viscoelastic relaxation, or damage, or from synthetic update rules implemented in engineered systems.

Most work on directed aging and physical learning has focused on the quasistatic limit, although several recent studies have begun to extend these ideas to dynamical settings~\cite{kendall2021gradient,lopez2023self,pourcel2025lagrangian,pourcel2025learning,massar2025equilibrium,berneman2025equili}. Implementing such methods in passive materials requires local evolution rules that are compatible with natural physical processes. Ideally, the rules should not require time-reversed dynamics or simultaneous access to two system states, as is common in contrastive learning methods. Here, we specialize in passive overdamped systems and derive tuning rules that are tailored for these systems.

Our approach is based on the Rayleighian~\cite{rayleigh1873some}, a quantity that depends on both forces and velocities and is instantaneously extremized by overdamped dynamics. Although the complete trajectory does not generally extremize the time-integrated Rayleighian, its instantaneous extremization provides a natural foundation for constructing approximate learning and adaptation schemes.

We develop two such schemes. The first is a directed-aging method that modifies microscopic parameters to create a soft mode in dissipation space, thereby steering the system's dynamical response. The second extends ideas from contrastive rules to the Rayleighian and produces local update rules that reduce a prescribed dynamical cost function.

We test both approaches numerically in networks of Maxwell viscoelastic elements, which respond elastically at short times and fluid-like at long times. We train their finite-rate response and demonstrate control over several dynamical properties. First, we tune the viscous Poisson ratio, which characterizes the response at slow actuation rates. We then train local input-output behaviors in which actuation at several source nodes produces a prescribed displacement at a distant target. Because the elastic and viscous responses are governed by distinct microscopic parameters the spring constants and dissipation coefficients, respectively the fast elastic response and the slow viscous response can be tuned independently. This separation enables programmable rate-dependent functionality.
{ We also shows that one can tune periodic steady states in a minimal model. }

Overall, our results show that systems with many overdamped degrees of freedom can be trained to exhibit prescribed dynamical responses. The methods introduced here may apply broadly to soft and biological materials and, more generally, to active or living systems in which local adaptation rules are naturally coupled to the dynamics.

\section{Overdamped dynamics \& The Rayleighian}

We consider arbitrary overdamped equations of motion where the velocities are linearly dependent on the forces, 
\begin{equation}
\sum_{j}\eta_{ij}\dot{x}_{j}=f_{i}\left(\left\{ x\right\} \right).
\end{equation}
Here, $x_i$ are the particle positions, $f_i$ are internal or external forces, and $\eta_{ij}$ is the dissipation matrix,
which is symmetric and positive-definite. Symmetry is due to Onsager's
reciprocal theorem~\cite{onsager1931reciprocal,doi2021onsager} and
positivity ensures that the system cannot spontaneously generate energy. 

Methods for controlling the response of elastic or other physical
systems often rely on the state being a minimum of some function.
For example, in quasistatics, the energy is minimized, and for resistor
networks, the dissipated power is minimized. For overdamped dynamics, the
state of the system can be defined in either the position of each
particle or the velocities, while also specifying the initial conditions.
The Rayleighian~\cite{rayleigh1873some,doi2021onsager} is a function whose extremum
with respect to the velocities yields the equation of motion. It is defined in
terms of the dissipation and the rate of change in potential energy, 
\begin{equation}
R=\frac{1}{2}\sum_{i,j}\eta_{ij}\dot{x}_{i}\dot{x}_{j}-\sum_{i}f_{i}\left(\left\{ x\right\} \right) \dot{x}_{i}.\label{eq:Rey}
\end{equation}
As noted, requiring that $\frac{\partial R}{\partial\dot{x}_{i}}=0$ yields the equations of motion. Since the dissipation matrix is positive-definite, the Rayleighian is minimized. 

Eq. \ref{eq:Rey} is remarkably similar to elastic systems, where the energy, in the linear regime, can be expressed in terms of the displacements, 
\begin{equation}
U=\frac{1}{2}\sum_{i,j}x_{i}H_{ij}x_{j}-\sum_{i}f_{i}^{ext}x_{i}.
\end{equation}
 where $H_{ij}=\frac{\partial^{2}U}{\partial x_{i}\partial x_{j}}$
is the Hessian and $f_{i}^{ext}$ is an external force. The Hessian
is analogous to the dissipation matrix, the velocities correspond
to the displacements and the internal forces correspond to the
external forces. 

While the momentary evolution is governed by the Rayleighian, the  trajectory \emph{does not} extremize the integrated Rayleighian,
\begin{equation}
R_{traj}=\int_{\tau_{i}}^{\tau_{f}}d\tau\,R\left(\tau\right).
\end{equation}
To see this, consider the internal forces (and possibly the dissipation matrix), which depend on the particle positions. The positions implicitly depend on past velocities,
$x_{i}\left(\tau\right)=x_{i}\left(0\right)+\int_{0}^{\tau}d\tau'\,\dot{x}\left(\tau'\right)$ and taking the functional derivative with respect to the velocity at a given time yields additional terms. { This defines a temporal Greens function $\partial x_i(t)/\partial\dot{x}_j(t')$. We speculate that if this Greens function decays quickly the integrated Rayleighian approximately extremizes the entire trajectory}. Another limit where the integrated Rayleighian is minimized exactly is if the forces and the dissipation matrix are constant, and do not depend on positions.

\section{Model}

\begin{figure}
\includegraphics[scale=0.5]{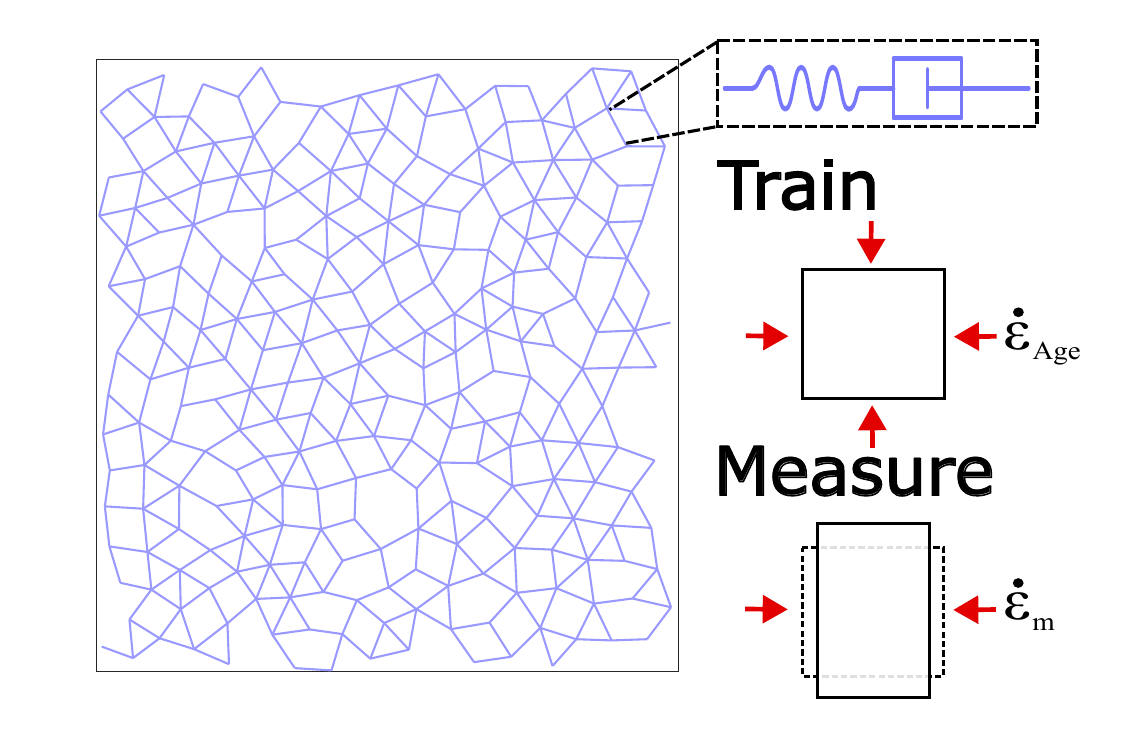}

\caption{An illustration of the model. On the left, an example of a disordered bonded network. Each bond is a Maxwell element, with a spring and dashpot in series. We train for a negative Poisson's ratio by uniformly compressing the system at a strain rate $\dot{\epsilon}_{Age}$. After training, the Poisson's ratio is measured by uniaxially straining along the x-axis at a strain rate $\dot{\epsilon}_{m}$ and measuring the transverse strain. \label{fig:Fig0}}
\end{figure}

For concreteness, we test our ideas on a specific model, a disordered network of 
 Maxwell viscoelastic elements, illustrated in Fig. \ref{fig:Fig0}~\cite{maxwell1867iv,lakes2009viscoelastic,zaccone2023general}. The network
is a two-dimensional disordered bonded network, where each bond is
composed of a spring and a dashpot in series. For convenience, the networks
are derived from amorphous packings of spheres where the coordination
number is easily tuned~\cite{ohern2003jamming}. The coordination number is the average number
of bonds per node, $Z=\frac{2N_{B}}{N}$, where $N_{B}$ is the number
of bonds and $N$ is the number of nodes. Here, we consider highly
coordinated networks with $Z=\frac{2N_{B}}{N}\approx4.52$, far from
the threshold needed for rigidity $Z_{c}=4$.

As noted, each bond is a spring and dashpot in series. The tension
on each bond is
\begin{equation}
t_{i}=k_{i}\left(\ell_{i}-\ell_{i,0}\right),\label{eq:tension}
\end{equation}
where $k_{i}$ denote the spring constants, $\ell_{i}$ are the bond
lengths and $\ell_{i,0}$ are the rest lengths. The dashpot accounts
for the change in rest length, assumed to be in proportion to the
tension on the bonds:
\begin{equation}
\gamma_{i}\frac{d}{d \tau}\ell_{i,0}=t_{i}.\label{eq:dashpot}
\end{equation}
Here, $\gamma_{i}$ are the dissipation coefficients of the dashpots. 

In this paper, {we assume a separation of time scales between the evolution of the dashpots and the time scale to reach force balance. Namely, the dashpots are the slow dynamical variables and the positions of the nodes are determined by force balance condition which depends on the rest lengths (the slow variables). }
Time is discretized into small intervals, and at each step we compute the tensions, change the rest lengths in accordance with Eq. \ref{eq:dashpot}
and then minimize the energy to reach force balance. 

For this model, the Rayleighian is given by, 
\begin{equation}
R=\frac{1}{2}\sum_{i}\gamma_{i}\left(\frac{d\ell_{i,0}}{d \tau}\right)^{2}-\frac{d\ell_{i,0}}{d \tau }k_{i}\left(\ell_{i}-\ell_{i,0}\right)
\end{equation}
where $\ell_{i}$ is determined by the force-balance requirement.

\section{Training rules}

Our goal in this paper is to control trajectories of certain target
degrees of freedom. These could be the positions of a specific set of
target particles or strain of the entire system, which we later discuss.
To this end, we strain the system at a constant rate and modify $\gamma_{i}$
to attain a desired behavior. Based on the analogy to quasistatic
mechanics, we employ two methods for adjusting the dissipation coefficients:\\
\textbf{(1) Directed aging approach:} which assumes that internal
parameters vary over time due to natural processes, such as aging, plasticity, or damage~\cite{pashine2019directed}. This microscopic change to material parameters translates to a change to material properties, including the response to external forcing.

To illustrate how this can be used to tune material properties, we review a rule that was used to tune elastic properties~\cite{pashine2019directed}.  Consider an elastic system where a bond accumulates damage and weakens over time. That is, we assume that a bond's stiffness decreases at a rate that grows with the tension on that bond, 
\begin{equation}
\frac{dk_{i}}{d\tau}=-r_{k}k_{i}\left(\ell_{i}-\ell_{0.i}\right)^{2}.\label{eq:k_model-1}
\end{equation}
Here, $r_k$ is a rate constant. The tension on a bond depends on the deformation, and therefore, different deformations result in different changes to material properties. For example, under compression, the bonds that counter this deformation weaken at a faster rate, resulting in the decrease of the bulk modulus. This also has the effect of decreasing the Poisson's ratio, since it
is a decreasing function of the ratio of the shear modulus to the
bulk modulus, $G/B$~\cite{pashine2019directed}
(in two dimensions it is given by $\nu_{2d}=\frac{B-G}{B+G}$) .
Straining under shear, for example, yields a small shear modulus.

For the time being, we do not alter the spring stiffness but rather define a rule that is motivated by Eq. ~\ref{eq:k_model-1}.
To control the viscous evolution of the Maxwell material, we alter
the damping coefficients. Based on the analogy between the Rayleighian { and}
 the energy, we assume that the dissipation coefficients degrade
at a rate that depends on the creep (plastic strain), 
\begin{equation}
\frac{d\gamma_{i}}{d\tau}=-\gamma_{i}r_{Age}\left(\frac{d}{d\tau}\ell_{i,0}\right)^{2}.\label{eq:TrainingRule}
\end{equation}
This evolution is very similar to Eq. \ref{eq:k_model-1}. Rather
than creating a soft direction in energy space, this rule creates a soft
direction in the dissipation space. { The soft direction is along the small eigenmodes of the dissipation matrix}. We also include the prefactor
$\gamma_{i}$ which insures that the $\gamma_{i}$ does not become
negative. 

\noindent \textbf{(2)Contrastive approach:} Contrastive methods ~\cite{ackley1985learning,movellan1991contrastive} learn by comparing a system's free response with its response under a target-dependent perturbation. Parameter updates are determined from the difference between the two states. Equilibrium propagation~\cite{scellier2017equilibrium,stern2021supervised} is an instance of a contrastive method that computes the gradient of an arbitrary cost function exactly. The condition of this theorem is that the system extremizes an energy function, which need not be the physical energy. For the sake of discussion, we focus on quasistatics where the potential energy is minimal. 

Equilibrium propagation is a supervised learning rule in which a supervisory signal `nudges' the system toward the desired behavior. This effect is incorporated through an additional term in the potential energy, given by the cost function weighted by~$\beta$:
\begin{equation}
E\rightarrow E+\beta C.
\end{equation}
Since the cost is minimized at the desired behavior, the additional force  `nudges' the system towards that desired behavior. The gradient of the cost function is given by,
\begin{equation}
\frac{\partial C}{\partial\gamma_{i}}\approx\frac{1}{\beta}\left[\frac{\partial E^{c}}{\partial\gamma_{i}}-\frac{\partial E^{f}}{\partial\gamma_{i}}\right].
\end{equation}
The superscript $c$ denotes the \emph{clamped state} where the system is
nudged towards the desired behavior ($\beta>0$) due extra potential
term. The superscript $f$ denotes the \emph{free state} where there is no additional
forcing, $\beta=0$. This becomes an equality in the limit of $\beta\rightarrow0$.

To employ this equilibrium propagation in dynamics, we substitute the energy by the Rayleighian,  
\begin{equation}
\frac{\partial C}{\partial\gamma_{i}}\approx\frac{1}{2 \beta}\left[\left(\frac{d}{d\tau}\ell_{i,0}^{c}\right)^{2}-\left(\frac{d}{d\tau}\ell_{i,0}^{f}\right)^{2}\right].\label{eq:EqProp_gamma}
\end{equation}
Thus, the gradient of a cost function is estimated by the difference in the square of the rate of plastic strain of the dashpot. Note that in comparison
to Eq. \ref{eq:TrainingRule}, there is an extra term corresponding
to the free state. Next, we assume that $\gamma_i$ evolves through gradient descent 
\begin{equation}
\frac{d\gamma_{i}}{d\tau}\propto-\frac{\partial C}{\partial\gamma_{i}},
\end{equation}
where the gradient is estimated using the approximate equilibrium propagation rule.  In this paper, we use the terms contrastive rule and equilibrium propagation interchangeably.

\section{Numerical Simulations}

\subsection{Directed aging for viscous responses}

We begin by training viscous responses using the directed aging
rule, Eq. \ref{eq:TrainingRule}. We aim to reduce the Poisson's ratio to negative values, which is uncommon in nature.
We train the system by repeatedly compressing at a constant strain
rate, $\dot{\epsilon}_{Age}$, while allowing the dissipation coefficients
to evolve through Eq. \ref{eq:TrainingRule}. At each cycle, we reset
the system to the same initial conditions. After training, we measure
the response to an applied uniaxial strain along the x-axis at a constant
rate and measure the resulting strain along the y-axis. Recall, Poisson's ratio is defined as the negative ratio of the strain along
the y-axis and the x-axis, $\nu=-\frac{\epsilon_{y}}{\epsilon_{x}}.$

\begin{figure}
\includegraphics[scale=0.475]{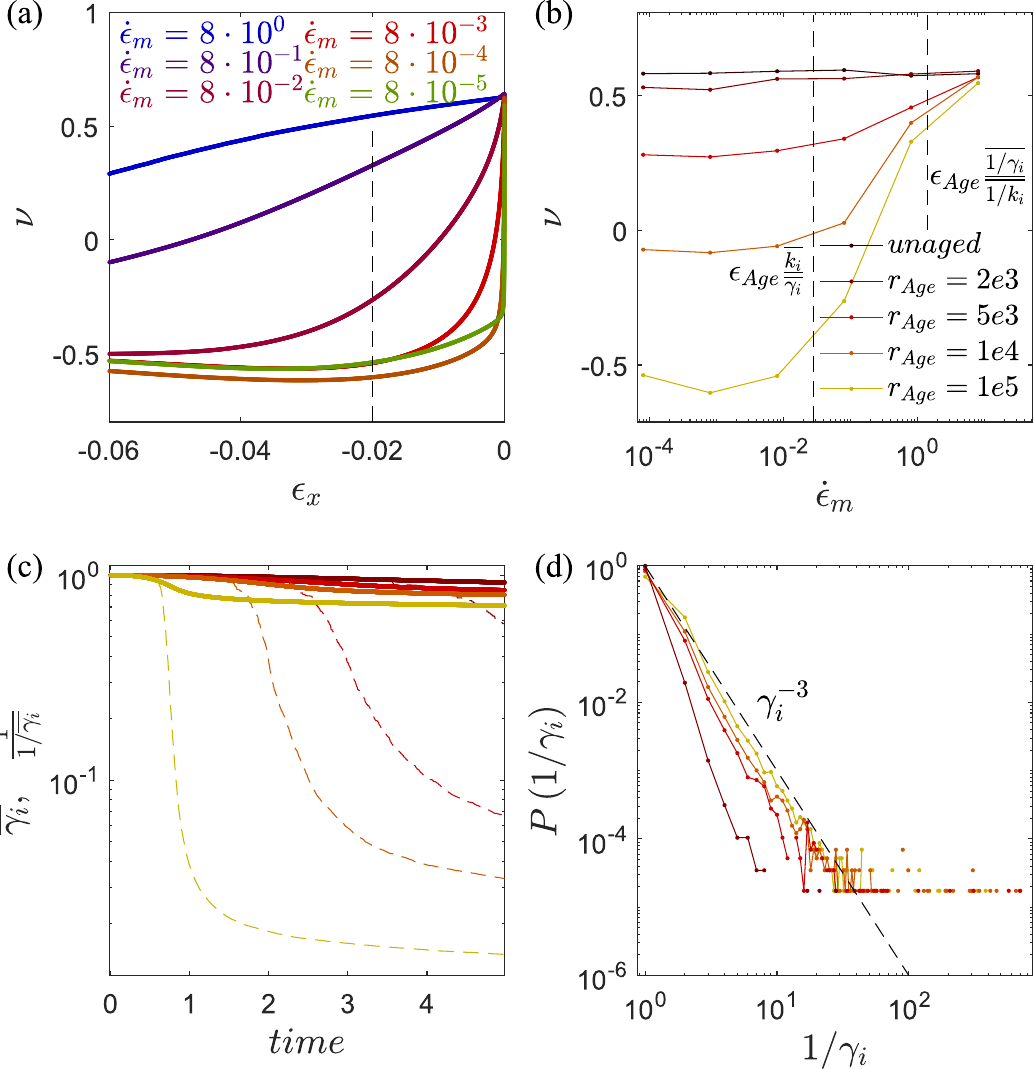}

\caption{Tuning the Poisson's ratio under compression using directed aging. (a) The Poisson's ratio as a function of strain for different actuation
rates. { The dashed line indicated $\epsilon_{Age}$} (b) The Poisson's ratio at the training strain as a function
of actuation rate, $\dot{\epsilon}_m$. The results are obtained for different training
rates. (c) The mean $\gamma_i$ { (solid lines)} and the inverse of the mean $1/\gamma_i$ { (dashed lines)} as a function of training time. The latter is dominated by a few bonds with a small $\gamma_i$.  (d) The distribution of $1/\gamma_i$ after training approaches a power-law distribution.  Parameters: $N=512$, $\dot{\epsilon}_{Age}\gamma_{i}\left(0\right)/k=8\cdot10^{-4}$,
$\epsilon_{Age}=0.02$ and number of training cycles is $10^5$. In (a): $r_{Age}=10^{5}$.\label{Fig1}}
\end{figure}

The Poisson's ratio after training is shown in Fig. \ref{Fig1}(a) at different deformation
rates, $\dot{\epsilon}_{m}$, measured under uniaxial compression. Note that the strain rate at which the response is measured, $\dot{\epsilon}_{m}$, is generally different from the training strain rate, $\dot{\epsilon}_{Age}$.  At fast deformation rates,
the response is dominated by the elastic response, which for this
ensemble of networks has a positive Poisson's ratio. At slow deformation
rates the Poisson's ratio depends on the strain. For small strains
(short times) the deformation is elastic, and the Poisson's ratio is
positive. At larger strains, where the viscous relaxation dominates,
Poisson's ratio becomes negative. The crossover between the elastic
and viscous becomes sharper, and in the limit of vanishing deformation
rate, there is a discontinuity in Poisson's ratio at zero strain. 

In Fig. \ref{Fig1}(b) we plot the Poisson's ratio at the training
strain $\epsilon_{Age}=0.02$ { (see table 1 for definition of the parameters)}, for different values of aging rates, $r_{Age}$.
The response of the system that was not aged ($r_{Age}=0)$, depends
very weakly on the strain rate. After training, for large enough aging
rates, Poisson's ratio crosses over from { positive} values at small strain
rates to { negative} values at large strain rates. 

We next ask what sets the strain rate of the crossover
between the two regimes. For a single dashpot, there is a characteristic strain rate, $\dot{\epsilon}=\epsilon_{Age}$$\frac{k_{i}}{\gamma_{i}}$.
For a network of springs, relaxation depends on the interactions of the bonds, which can be solved numerically. There are two simple cases for which the collective evolution can be computed: arranging the bonds
either in series or in parallel. Both cases map onto the single bond
case with an effective characteristic strain rate, $\epsilon_{Age}\frac{\overline{k_{i}}}{\overline{\gamma_{i}}}$
when in parallel and $\epsilon_{Age}\frac{\overline{1/\gamma_{i}}}{\overline{1/k_{i}}}$ when the bonds are in series. Here, the bar denotes the average over
all springs or dashpots, i.e., $\overline{k_{i}}=\frac{1}{N}\sum_{i}k_{i}$. These two values are indicated in Fig. \ref{Fig1}(b) by the dashed lines. The crossover occurs between these two values. 

We note that $\overline{\gamma_{i}^{-1}}$ and $(\overline{\gamma_{i}})^{-1}$
may differ by many orders of magnitude, since the first is dominated
by the small values, while the second by the large values. Fig. \ref{Fig1}(c)
compares $\overline{\gamma_{i}}$ and $1/\overline{\gamma_{i}^{-1}}$
as a function of time during training. While $\overline{\gamma_{i}}$ changes very little,
the change to $1/\overline{\gamma_{i}^{-1}}$ is significant.
We also plot in Fig. \ref{Fig1}(d) the distribution of $1/\gamma_{i}$.
The distribution for the large $r_{Age}$ approximately is power-law,
scaling as $\left(\frac{1}{\gamma_{i}}\right)^{-3}$. { Bonds with small $\gamma_i$ do not retard the motion and therefore the strain is larger along these bonds.  }

The probability
of $\gamma_{i}$ at small values can be estimated through a change
in variables, yielding $P\left(\gamma\right)\propto\gamma$. Namely, the distribution
vanishes linearly at small values. Power laws are often associated
with criticality. In the appendix \ref{sec:Appendix}, we provide a
one-dimensional model that yields this scaling. The aging process
is self-accelerated (positive feedback). Assuming the tension, $t$,
is fixed, combining the aging rule with the evolution of a dashpot yields
, $\frac{d\gamma_{i}}{d\tau}=-\frac{r_{Age}}{\gamma_{i}}t^{2}$. Namely,
bonds with small $\gamma_{i}$ decrease at a faster rate. This could
yield spatial localization of small $\gamma_{i}$, in a crack-like
fashion, similar to that occurring in elastic systems~\cite{goodrich2015principle}.
 
\begin{table}[h]
\centering
\begin{tabular}{l l}
\hline
Parameter & Definition  \\
\hline
$\epsilon_{Age}$ & Terminal training strain \\
$\dot{\epsilon}_{Age}$ & Training strain rate \\
$\dot{\epsilon}_{m}$  & Measuring strain rate \\
$\epsilon_{T}$  &  Strain on target \\
$\epsilon_{S}$  &  Strain on source \\
$\epsilon_{D}$  & Desired strain \\
$\gamma_i$  & Dissipation coefficient \\
$\ell_i$  & Bond length \\
$\ell_{i,0}$  & Bond rest length \\
$k_i$  & Spring constant \\
$\beta$  & Contrastive parameter \\
$r_{Age}$  & Aging rate \\
$t_{i}$  &  Tension on bond \\
\hline
\end{tabular}
\caption{Definition of important parameters}
\label{tab:example}
\end{table}

\begin{figure}
\includegraphics[scale=0.5]{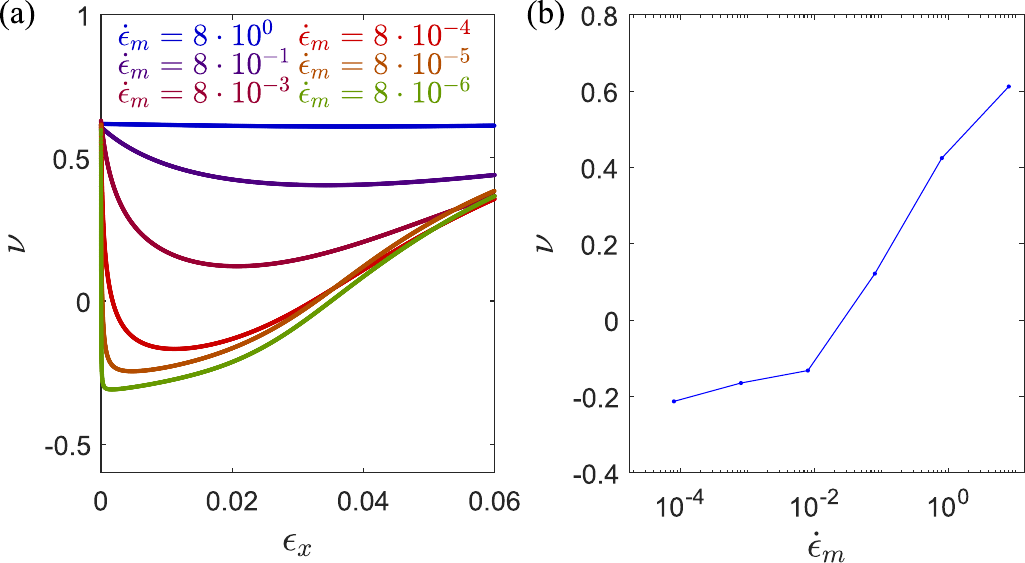}

\caption{Tuning the Poisson's ratio under expansion using directed aging. (a) The Poisson's ratio as a function of strain (expansion) for different actuation
rates. (b) The Poisson's ratio at the training strain as a function
of actuation rate. The results are obtained for different training
rates. Parameters: $N=512$, $\dot{\epsilon}_{Age}\gamma_{i}\left(0\right)/k=4\cdot10^{-3}$,
$\epsilon_{Age}=0.02$ and number of training cycles is $10^5$. \label{Fig2_expansion}}
\end{figure}

Next, we test the generality of our results and train auxetic responses
(negative Poisson's ratio) under expansion. We follow the exact procedure
as above. The results, shown
in Fig. \ref{Fig2_expansion}, are similar to those above. Poisson's
ratio at fast rates is positive, while at slow rates it can be negative.
Interestingly, there is a stronger dependence on the applied strain.
The Poisson's ratio as a function of strain first decreases, reaches
a minimum, and then increases. {  The minimum appears to  be on the scale of the training strain and is more pronounced than the case of compression.  The minimum is determined by the competition of several effects:  the transition from elastic to viscous response which occurs above a crossover strain, discussed above, and a memory of training strain. Interestingly at large enough strains the Poisson's ratio remains positive, indicating that training affects mostly strains near $\epsilon_{Age}$.  }
We also show in Fig. \ref{Fig2_expansion}(b) the Poisson's
ratio at the training strain as a function of the { measuring} strain rate. Similarly,
to the case of compression, there is a crossover between the quick
elastic with a positive Poisson's ratio and the slow viscous responses
with a negative Poisson's ratio. 

\subsection{Directed aging for local responses}

\begin{figure}
\includegraphics[scale=0.5]{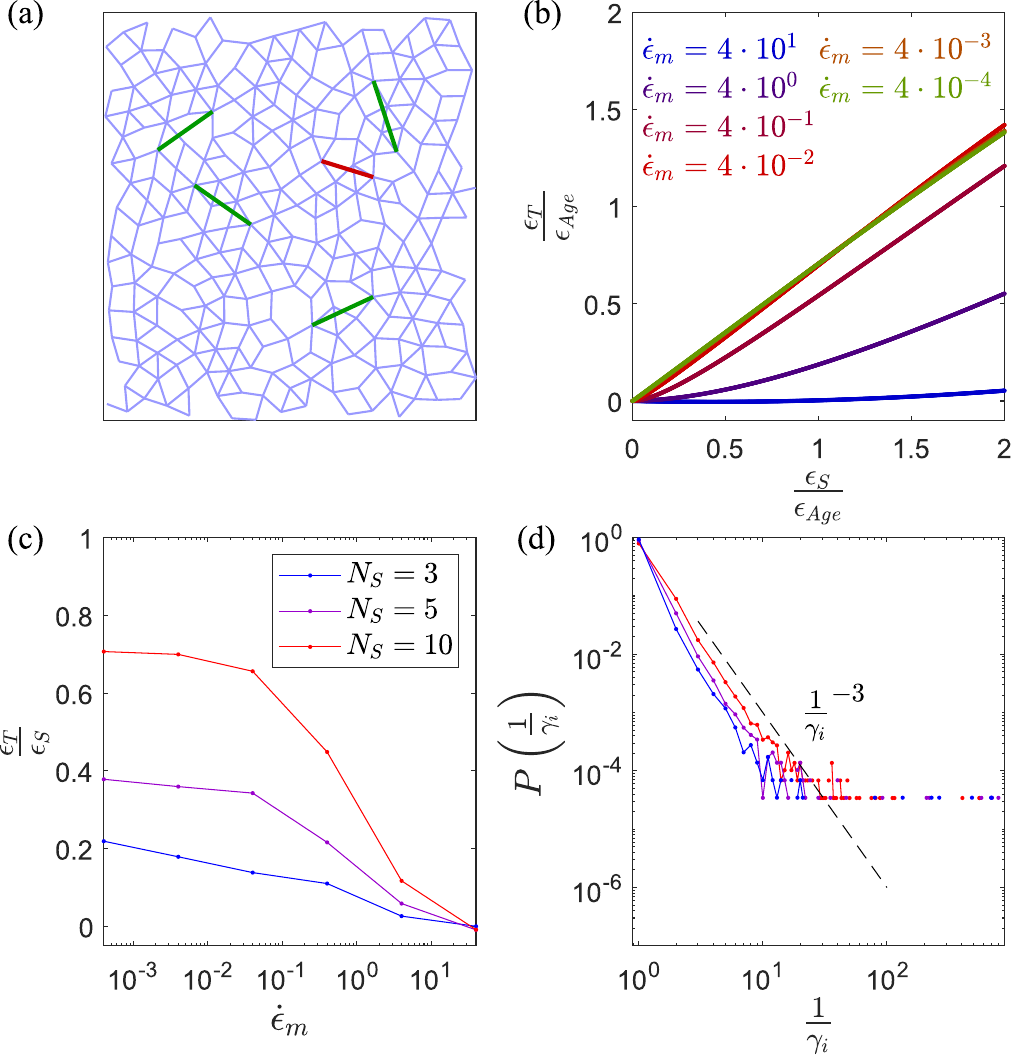}

\caption{Training for a local response using directed aging. (a) An illustration of the network with four sources (green)
and a single target (red). (b) The strain on the target as a function of
the strain on the source. (c) The average ratio of the strain of the
target to the source, as a function of actuation rate for different numbers of sources. (d) The distribution of the inverse dissipation parameters
appears to be power-law, similarly to the results for the Poisson's
ratio. Parameters: $N=256$, $\dot{\epsilon}_{Age}\gamma_{i}\left(0\right)/k=2\cdot10^{-3}$, 
 $\epsilon_{Age}=0.02$ and number of training cycles is $10^5$. In (b) $N_{S}=10$. \label{fig:Loca}}

\end{figure}

We also explore the possibility of training local responses. We consider
$N_{S}$ source sites that are actuated with the goal of inducing
a strain on a single target site. Both the sources and the target are a
pair of nearby nodes that are chosen randomly, and the strain is defined
as the fractional change in the distance between the nodes. Fig. \ref{fig:Loca}(a)
shows an example of the network with the source sites indicated
in green and the single target is indicated in red. We train at a
strain amplitude of $\epsilon_{Age}=0.02$ for both the sources and
the target; the applied strain is either compression or extension,
chosen with equal probability. 

Fig. \ref{fig:Loca}(b) shows the response at different strain rates
to an applied strain on the source sites, $\epsilon_S$. At fast rates, the response
is very small and, on average, vanishes. At slow rates, the response
approximately follows the desired strain, and the response persists beyond the training
strain. Fig. \ref{fig:Loca}(c) shows the strain on the target at
the training strain $\epsilon_{Age}$, for different numbers of source
sites. The response is the largest when there are sufficient number
of source sites. This is consistent with previous findings for other
training methods~\cite{hexner2020periodic}. The difficulty in training
local responses is that the elastic (or viscous) Green's function for
an applied local force decays quickly with distance from that site.
As a result, it is difficult to couple distant sites. 

Lastly, we consider the distribution of $1/\gamma_{i}$, shown in Fig.
\ref{fig:Loca}(d). Also here, the distribution scales approximately
as $\left(\frac{1}{\gamma_{i}}\right)^{-3}$. This suggests that this
distribution is universal for this training method. 

\subsection{Multi-functional responses }

Up till now, we have focused on manipulating the slow viscous response, which depends on the dissipation coefficients, $\gamma_{i}$. As noted at fast actuation rates, the response depends on the spring constants, $k_{i}$. Since the fast and slow response depends on different sets of parameters, this offers the possibility of modifying the elastic and dissipative responses independently. 

To explore the feasibility of tuning these two limits, we train the
system using two different training rules. For the sake of this example,
we train for a negative Poisson's ratio at fast rates and a positive
Poisson's ratio at slow rates. This is different from above (see Fig.
\ref{Fig1}(b)) where $\nu>0$ for fast deformations and $\nu<0$ for
slow deformations. The elastic response is altered by decreasing the
spring constants in proportion to the energy on a bond, as defined
by Eq. \ref{eq:k_model-1}. During training for the elastic response,
we compress the system isotropically, { as to decrease the resistance to compression}.

After training the elastic response, we then train the slow viscous response by altering the relaxation times
of the dashpots in Eq. \ref{eq:TrainingRule}. To increase the Poisson's
ratio, we train with a pure shear deformation { which decreases the resistance to shear}. We note that the two rules
do not interfere with each other. Training at fast rates generates large
stresses and therefore affects only the spring constants. Training
at slow rates is dominated by creep, altering the dissipation coefficients. 
{ Training independently these two processes assumes a separation in time scales between these two aging modes.}

\begin{figure}
\includegraphics[scale=0.48]{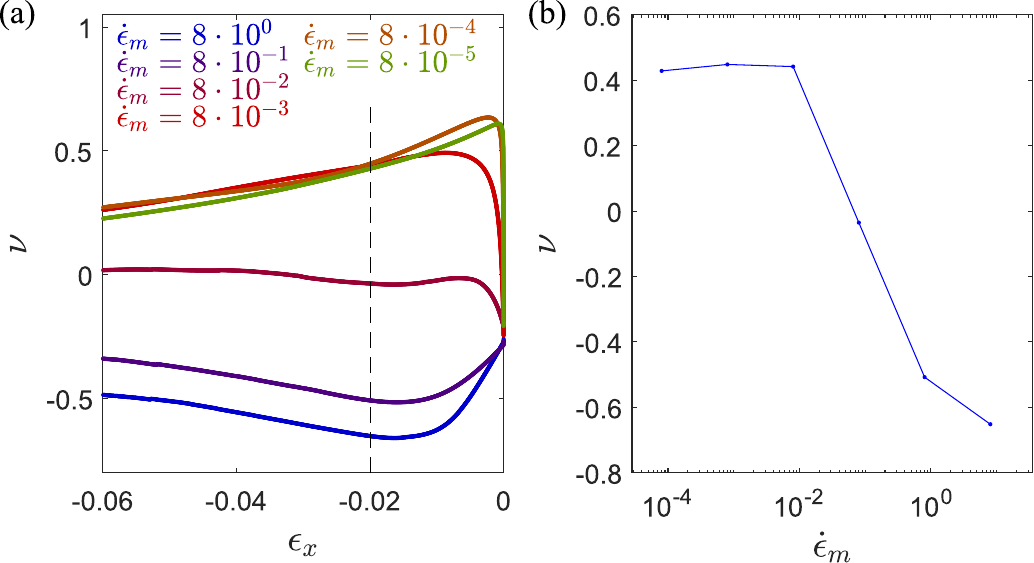}

\caption{Tuning both the elastic and viscous response using directed aging. (a) The Poisson's ratio as a function of strain for different actuation
rates. (b) The Poisson's ratio at the training strain as a function
of actuation rate. The results are obtained for different training
rates. Simulation parameters: $r_{Age}=5\cdot10^{5},$ $\epsilon_{Age}=-0.02$,
$N=512$, $\dot{\epsilon}_{train}=5$, $\dot{\epsilon}_{train\,k}r_{k}=0.01,$$\gamma_{0}=1$,
$k_{0}=1$ and number of training cycles is $10^5$.  \label{fig:multi_func}}
\end{figure}

Fig. \ref{fig:multi_func}(a) shows the Poisson's ratio as a function
of applied uniaxial strain. At fast actuation rates, the response has
$\nu<0$, while at slow rates $\nu>0$ as the system was trained for. Fig.
\ref{fig:multi_func} (b) shows $\nu$ at the training strain $\epsilon_{Age}=-0.02$,
as a function of the {measuring} strain rate $\dot{\epsilon}_{m}$. As shown,
there is a crossover between positive and negative Poisson's ratios as a function of actuation rate. 

\subsection{Equilibrium propagation for local responses}

Here we train local responses using the contrastive rule
introduced in Eq. \ref{eq:EqProp_gamma}. As before, we consider several
sources and a single target site, which corresponds to pairs of nearby
nodes. These are trained with the goal of attaining a desired strain
on the target when the sources are strained at a constant rate. We
iteratively vary the dissipation coefficients using an approximate gradient descent { scheme where the gradient is estimated using the approximate  contrastive rule}. In each cycle we begin from the same initial conditions. In
each iteration, we strain the system twice, first, in the free state
and then in the clamped state. In the free state, we do not apply any
forcing to the target, while in the clamped state, we nudge the targets.
The small nudging force acts to adjust the distance between the target
nodes toward their desired value. Here we choose to take the desired
strain to be $\epsilon_{T}=\epsilon_{Age}\left(\tau/T\right)^{2}$, where $\tau$ is time and 
$T$ is a time constant. We choose the quadratic form since it
is more consistent with the viscous response. We also limit the maximal
and minimal dissipation coefficients.

\begin{figure}
\begin{centering}
\includegraphics[scale=0.6]{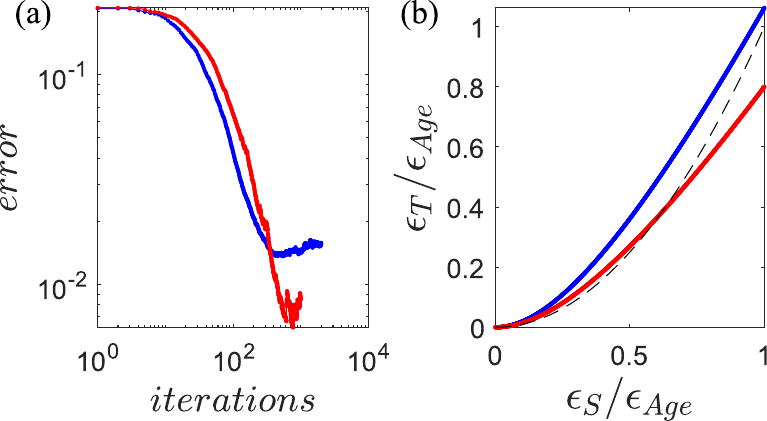}
\par\end{centering}
\caption{Tuning a localized allostery-like response using the contrastive rule. (a) The
error as a function of the number of iterations. (b) The strain on
the target as a function of the strain on the source. Blue: We nudge
the system with a force that is proportional to the difference in the
velocity and the desired velocity $\left(\dot{\epsilon}_{T}-\dot{\epsilon}_{D}\right)$.
Red: The nudging depends on the distance between the nodes of the target
and their desired value. The dashed line corresponds to the desired strain.
Here, $N=256$, actuation time normalized by dissipation coefficient
$\frac{\epsilon_{Age}k}{\dot{\epsilon}_{Age}\gamma}=0.04$, $\epsilon_{Age}=0.02$,
$N_{S}=5$ and $\beta=0.001$. \label{fig:EqProp}}
\end{figure}

In Fig. \ref{fig:EqProp}(a), we show the error as a function of the number of iterations. The error is defined as the integrated square difference of the target strain rate and the desired strain rate $\frac{1}{T\epsilon_{Age}^{2}}$$\int_{0}^{T}d\tau\left(\dot{\epsilon}_{T}-\dot{\epsilon}_{D}\right)^{2}$. We consider two variations
of the nudging force: in the blue curve, the force is proportional to
the difference between the velocity in the free state and the desired
velocity, while in the red curve, the nudging depends on the difference
between the target strain and the desired strain. The nudging force is only approximate
and is not derived from the Rayleighian, which is in itself only approximately
extremized. For both cases, the error decreases by over an order of
magnitude. Fig. \ref{fig:EqProp} shows the strain on the target as
a function of the source strain. For both nudging forces, the strain
is near the desired values, indicated by the dashed line.

{
\subsection{Equilibrium propagation for periodic steady states}

We also test whether equilibrium propagation can be used to train responses that are intrinsically dynamical, such as periodic steady state. We consider a minimal overdamped system driven periodically composed of three nodes and two bonds in one dimension. A single output coordinate $x(t)$ is connected by Kelvin-Voigt elements to a fixed node and to a driven node with prescribed motion
\begin{equation}
x_{drv}(t)=A_{drv}\cos(\omega t).
\end{equation}
 We assume that the spring constants \(k_1,k_2\) are fixed, while the viscous coefficients \(\gamma_1,\gamma_2\) are trainable. The free dynamics are
\begin{equation}
(\gamma_1+\gamma_2)\dot{x}
+(k_1+k_2)x
= k_2x_{drv}+\gamma_2\dot{x}_{drv}.    
\end{equation}

After transients decay, the system approaches a periodic steady state whose amplitude and phase depend on the dissipative parameters. We train the output to follow $x_T(t)=A_T\cos(\omega t-\phi_T)$.

As above, the system is first evolved freely and then weakly nudged toward the target by a force $\beta[x_T(t)-x(t)]$. The two damping coefficients are updated from the contrast between the local dissipation in the nudged and free trajectories,

\begin{align}
 \Delta\gamma_1 &\propto -\frac{1}{2\beta} \left[
(\dot{x}^{\,\beta})^2-(\dot{x}^{\,0})^2\right], \\
    \Delta\gamma_2   &\propto -\frac{1}{2\beta}
\left[ (\dot{x}^{\,\beta}-\dot{x}_{drv})^2-(\dot{x}^{\,0}-\dot{x}_{drv})^2 \right].
\end{align}
By adapting only the local dissipation, we train the system to tune both the amplitude and phase of its periodic steady-state response.

Fig. \ref{fig:periodic} shows numerical results for training a desired phase shift and amplitude relative to the driven node. Since the model has only two learning degrees of freedom the range of allowed phases and amplitudes is limited and are chosen accordingly. Panel (a) shows that the error decreases by two ordered of magnitude. Panel (b) shows the displacement of the driven node and of the target before and after training. After training the displacement of the target matches well the desired value. Choosing other viable parameters shows similar convergence.

\begin{figure}
\includegraphics[scale=0.48]{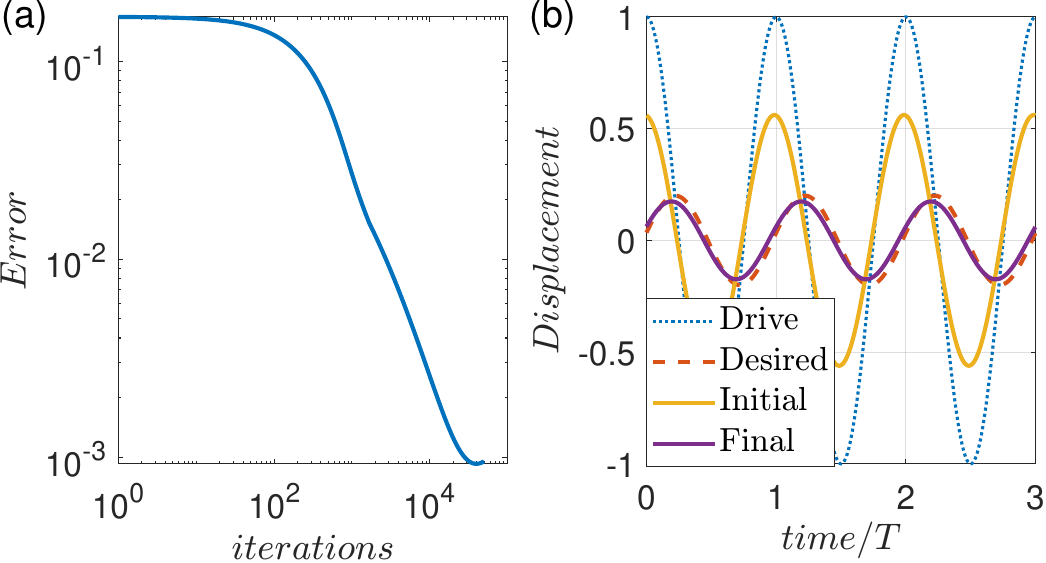}

\caption{{ Training periodic motion in a system composed of two Voigt bonds where a single node driven periodically. Our goal is to control the amplitude and phase of the target. (a) The mean square error as a function of iterations. (b) The displacement of the driven node, and the target before and after training. We also show the desired trajectory. Goal: $80^o$ phase shift and an amplitude of $0.2$ of the driving amplitude.} \label{fig:periodic}}
\end{figure}

}
\section{Conclusions }

In summary, we have shown that the overdamped dynamics of materials
can be tuned to allow desired behaviors. We have developed two rules
that are derived from the Rayleighian, whose extremum corresponds to the equations of motion. While this is exact at a given point in time,
for a trajectory, the integrated Rayleighian is not extremized, and
therefore the rules are only approximate. The structure of the Rayleighian
is very similar to the elastic energy, where the dissipation matrix
corresponds to the elastic Hessian. Based on this analogy, we have derived
a directed aging rule where the dashpots degrade as they evolve, creating
a soft direction in the dissipation space. A second rule is derived
through the equilibrium propagation framework. 

We then tested these ideas in a model consisting of a disordered
bonded network where each bond is a spring and dashpot in series.
This is the Maxwell model for viscoelastic materials, and its behavior
is rate-dependent. On short time scales (fast actuation rates), it
is elastic, while on long time scales (slow actuation), it undergoes
dissipative flow. The rules are applied to the dissipation coefficients,
allowing for tuning the slow relaxation processes. This allows for modifying
the Poisson's ratio as well as the local responses that arise due
to flow, rather than elasticity. We then studied the feasibility of
tuning both the elastic and viscous properties simultaneously. Since
the elastic properties depend on the spring constants, they can be modified
independently of the dissipation coefficients, which determine the
viscous response. As a result, we can modify the rate-dependent
response, both for fast and slow deformations. { We also showed that the Rayleighian formalism allows to train periodic dynamics using equilibrium propagation.  }

{ 
Recently, there has been growing interest in extending physical learning and directed-aging rules to dynamical systems~\cite{mandal2024learning,stern2022physical,massar2025equilibrium,berneman2025equili,stern2026contrastive,scurria2026equilibrium,floyd2026learning,li2026situ}. However, a general, exact, and fully local learning rule for dynamical systems that break time-reversal symmetry remains lacking. Several approaches have been proposed, including rules that are exact only for linear systems~\cite{berneman2025equili}, approximate schemes that nudge the system along its forward trajectory~\cite{stern2026contrastive}, and the rules we have developed based on instantaneous minimization of the Rayleighian. Our approach is restricted to passive reciprocal systems. For materials, we view the directed-aging formulation as particularly relevant, since it does not require comparing two separate measurements or measuring a time reversed trajectory~\cite{berneman2025equili,li2026situ}. The regime of validity of each of the rules is still an open question. }

While the current paper has focused on a specific model, controlling
the overdamped dynamics could be important for a large set of systems.
In materials, this could allow for manipulating the dynamic moduli of
viscoelastic solids, the flow of suspensions, and electronic transport.
In living systems, on the cellular or molecular scale, dynamics are
often overdamped, and the current ideas could be relevant there. Lastly,
active systems are also often modeled with first-order equations of
motion. Pursuing the line of research we presented, perhaps we could
make the methods precise and find applications in both passive and
active systems. 
\begin{acknowledgments}
    This work was supported by the Israel Science Foundation (grant 2385/20) and the Alon Fellowship.
\end{acknowledgments}

\section{Appendix: distribution of $\frac{1}{\gamma_{i}}$ in one dimension\label{sec:Appendix}}

In this appendix, we compute the distribution of $\gamma_{i}$ for Maxwell elements arranged in series. Since the Maxwell elements are arranged in series, they all experience the same tension, $t$. Each dashpot evolves in
proportion to the tension,
\begin{equation}
\gamma_{i}\frac{d}{d\tau}\ell_{0,i}=t.
\end{equation}
We strain the system at a constant rate, which corresponds to increasing the tension at a constant rate $r$.
The tension relaxes due to the extension of the dashpots,
\begin{align}
\frac{dt}{d\tau} & =r+k_{eff}\sum\frac{\ell_{0,i}}{d\tau}\\
 & =r+k_{eff}t\sum\frac{1}{\gamma_{i}}
\end{align}
We will assume that the tension reaches a steady state and that it
remains constant. This neglects the effect of the change of $\gamma_{i}$ on the tension. The directed aging rule assumes that, 
\begin{align}
\frac{d\gamma_{i}}{d\tau} & =-\gamma_{i}\left(\frac{d\ell_{0,i}}{d\tau}\right)^{2}\\
 & =-\frac{t^{2}}{\gamma_{i}}.
\end{align}
In the last line, we used the relation between the tension and change
in rest length. Alternatively, in term of $\frac{1}{\gamma_{i}}$, 

\begin{equation}
\frac{d}{d\tau}\frac{1}{\gamma_{i}}=\frac{t^{2}}{\gamma_{i}^{3}}.
\end{equation}
The distribution of $\mu_{i}=\gamma_{i}^{-1}$ obeys the continuity
equation, 

\begin{equation}
\frac{\partial P\left(\mu\right)}{\partial t}=-\frac{\partial}{\partial\mu}\left(P\mu^{3}\right)
\end{equation}
This equation has no steady state, since $\gamma_{i}$ constantly
decreases. To estimate the distribution, we add a delta function $\delta\left(\mu-1\right)$
which injects bonds with $\gamma=1$. The steady state is given by
(for $\mu>1)$,  
\begin{equation}
P\left(\mu\right)=\frac{1}{\mu^{3}}.
\end{equation}
This is consistent with the simulations, despite them being in two
dimensions. 

\bibliographystyle{unsrt}
\bibliography{biblo}

\end{document}